\documentclass[12pt,preprint]{aastex}
\usepackage{lscape}

\newcommand{\bdv}[1]{\mbox{\boldmath$#1$}}

\def\rel{{\rm rel}}
\def\e{{\rm E}}
\def\au{{\rm AU}}
\def\muas{{\mu\rm as}}
\def\masyr{{\rm mas}\,{\rm yr^{-1}}}
\def\kms{{\rm km}\,{\rm s}^{-1}}
\def\kpc{{\rm kpc}}

\def\rel{{\rm rel}}
\def\bv{{\bf v}}

\def\e{{\rm E}}
\def\bpi{{\bdv{\pi}}}
\def\balpha{{\bdv{\alpha}}}

\def\bj{{\bf j}}

\begin{document}
\title{MOA 2003-BLG-37: A Bulge Jerk-Parallax Microlens Degeneracy}

\author{
B.-G. Park\altaffilmark{1},
D.L. DePoy\altaffilmark{2},
B.S. Gaudi\altaffilmark{3},
A. Gould\altaffilmark{2},
C. Han\altaffilmark{2,4},
and R.W. Pogge\altaffilmark{2} \\
(The $\mu$FUN Collaboration) \\ and \\
F. Abe\altaffilmark{5},
D.P. Bennett\altaffilmark{6},
I.A. Bond\altaffilmark{7},
S. Eguchi\altaffilmark{5},
Y. Furuta\altaffilmark{5},
J.B. Hearnshaw\altaffilmark{8},
K. Kamiya\altaffilmark{5},
P.M. Kilmartin\altaffilmark{8},
Y. Kurata\altaffilmark{5},
K. Masuda\altaffilmark{5},
Y. Matsubara\altaffilmark{5},
Y. Muraki\altaffilmark{5},
S. Noda\altaffilmark{9},
K. Okajima\altaffilmark{5},
N.J. Rattenbury\altaffilmark{10},
T. Sako\altaffilmark{5},
T. Sekiguchi\altaffilmark{5},
D.J. Sullivan\altaffilmark{11},
T. Sumi\altaffilmark{12},
P.J. Tristram\altaffilmark{10},
T. Yanagisawa\altaffilmark{13}, and
P.C.M. Yock\altaffilmark{10}\\
(The MOA Collaboration)}
\affil{
\altaffiltext{1}
{Korea Astronomy Observatory,
61-1, Whaam-Dong, Youseong-Gu, Daejeon 305-348, Korea; bgpark@boao.re.kr}
\altaffiltext{2}
{Department of Astronomy, The Ohio State University,
140 West 18th Avenue, Columbus, OH 43210, USA; depoy, gould, 
pogge@astronomy.ohio-state.edu}
\altaffiltext{3}
{Harvard-Smithsonian Center for Astrophysics, Cambridge, MA 02138, USA; 
sgaudi@cfa.harvard.edu}
\altaffiltext{4}
{Department of Physics, Institute for Basic Science Research,
Chungbuk National University, Chongju 361-763, Korea;
cheongho@astroph\-.chungbuk.ac.kr}
\altaffiltext{5}
{Solar-Terrestrial Environment Laboratory, Nagoya University,
Nagoya 464-8601, Japan; abe, furuta, kkamiya, kmasuda, kurata, 
muraki, okajima, sado, sako, sekiguchi, ymatsu@stelab.nagoya-u.ac.jp}
\altaffiltext{6}
{Department of Physics, Notre Dame University, Notre Dame, IN 46556, USA;
bennett@emu.phys.nd.edu}
\altaffiltext{7}
{Institute for Astronomy, University of Edinburgh, Edinburgh, EH9 3HJ, UK;
iab@roe.ac.uk}
\altaffiltext{8}
{Department of Physics and Astronomy, University of Canterbury,
Private Bag 4800, Christchurch, New Zealand;
john.hearnshaw, pam.kilmartin@canterbury.ac.nz}
\altaffiltext{9}
{National Astronomical Observatory of Japan, Tokyo, Japan;
sachi.t.noda@nao.ac.jp}
\altaffiltext{10}
{Department of Physics, University of Auckland, Auckland, New Zealand;
nrat001@phy.auckland.ac.nz, paulonieka@hotmail.com,
p.yock@auckland.ac.nz}
\altaffiltext{11}
{School of Chemical and Physical Sciences, Victoria University,
PO Box 600, Wellington, New Zealand; 
denis.sullivan@vuw.ac.nz}
\altaffiltext{12}
{Department of Astrophysical Sciences, Princeton University,
Princeton NJ 08544, USA; sumi@astro.princeton.edu }
\altaffiltext{13}
{National Aerospace Laboratory, Tokyo, Japan; tyanagi@nal.go.jp}
}

%\submitted{Submitted to The Astrophysical Journal}

\begin{abstract}

We analyze the Galactic bulge microlensing event MOA 2003-BLG-37.
Although the Einstein
timescale is relatively short, $t_\e=43\,$days, the lightcurve displays
deviations consistent with parallax effects due to the Earth's accelerated
motion.  We show that the $\chi^2$ surface has four distinct local minima
that are induced by the ``jerk-parallax'' degeneracy, with pairs
of solutions having projected Einstein radii, $\tilde r_\e=1.76\,\au$ and
$1.28\,\au$, respectively.
This is
the second event displaying such a degeneracy and the first toward
the Galactic bulge.  For both events, the jerk-parallax formalism
accurately describes the offsets between the different solutions,
giving hope that when extra solutions exist in future
events, they can easily be found.
However, the morphologies of the $\chi^2$
surfaces for the two events are quite different, implying that
much remains to be understood about this degeneracy.

\end{abstract}


\section{Introduction
\label{intro}}
	
	Microlens parallaxes are playing an increasingly important
role in the analysis of microlensing events.  The microlens parallax, 
$\bpi_\e$ is a vector, whose magnitude is the ratio of the size of
the Earth's orbit (1 AU) to the Einstein radius projected onto the
plane of the observer $(\tilde r_\e=\au/\pi_\e)$, and whose direction is that
of the lens-source relative motion.
 From simple geometrical considerations (e.g., \citealt{natural}),
\begin{equation}
\pi_\e = \sqrt{\pi_\rel\over\kappa M},\qquad
\theta_\e = \sqrt{\kappa M \pi_\rel},\qquad 
\kappa \equiv {4 G \over c^2\au} \simeq 8.1{{\rm mas}\over M_\odot},
\label{eqn:piedef}
\end{equation}
where $M$ is the mass of the lens, $\pi_\rel$ is the lens-source
relative parallax, and $\theta_\e$ is the angular Einstein radius.
Hence, if both $\pi_\e$ and $\theta_\e$ can be measured, one can
determine both $M$ and $\pi_\rel$.  While this has only been done
for one microlensing to date \citep{jin}, there are plans to apply
this technique to several hundred events using the 
{\it Space Interferometry Mission (SIM)} \citep{gssim}.  

	Moreover, even in cases for which $\theta_\e$ cannot be measured, 
the microlens parallax can give very important information about the
event.  When combined with the Einstein timescale, $t_\e$, it yields
$\tilde \bv$, the lens-source relative velocity projected onto the
observer plane,
\begin{equation}
\tilde \bv = {\bpi_\e\over \pi_\e^2}{\au\over t_\e}.
\label{eqn:tildev}
\end{equation}
By combining measurements of $\bpi_\e$ and $t_\e$ with known Galactic 
structure parameters,  \citet{O9932} and \citet{blackhole} 
were able to show that several
microlensing events seen toward the Galactic bulge are most probably
due to black holes.

	Microlens parallax measurements are possible whenever $\tilde r_\e$
can be compared to some ``standard ruler'' in the observer plane.  The two
logical possibilities are observing the event simultaneously from two
different locations and observing the event from an accelerated platform.
Since $\tilde r_\e$ is typically of order an AU, 
the most practical realizations 
of these ideas would be, respectively, to observe the event simultaneously
from a satellite in solar orbit \citep{refsdal66} and (for sufficiently long
events) to observe it from the Earth as it moves about the Sun \citep{gould92}.

	When \citet{refsdal66} first proposed obtaining microlens
parallaxes by combining space-based and ground-based observations
of microlensing events, he already noted that these measurements
would generically be subject to an ambiguity.  Using somewhat different
notation, he showed that,
\begin{equation}
\bpi_\e \equiv (\pi_{\e,\parallel},\pi_{\e,\perp})
= {\au\over d_{\rm sat}}(\Delta\tau,\Delta u_0)
\label{eqn:spacedegen}
\end{equation}
where $d_{\rm sat}$ and the parallel ($\parallel$) direction are defined by
the magnitude and direction of the 
Earth-satellite separation vector projected onto
the plane of the sky, $\Delta \tau = \Delta t_0/t_\e$ is the difference
in times of maximum, $t_0$, as seen from the Earth and satellite
(normalized to $t_\e$), and $\Delta u_0$ is the difference in impact
parameters.  While $\Delta\tau$ (and so $\pi_{\e,\parallel}$) is
unambiguously defined, 
\begin{equation}
\Delta u_0 = \pm |u_{0,\oplus}\pm u_{0,\rm sat}|,
\label{eqn:deltabeta}
\end{equation}
and so $\pi_{\e,\perp}$ is subject to a two-fold degeneracy in magnitude, 
depending on whether the event impact parameters $u_0$ register the lens
passing on the same $(-)$ or opposite $(+)$ sides of the lens as seen
from the Earth and satellite.  And $\pi_{\e,\perp}$ is subject to an additional
two-fold ambiguity in sign, depending on the relative orientations of
these separations.  See \citet{gould94} for a fuller discussion of
this degeneracy.

	Although no space-based parallaxes have ever
been measured, this degeneracy and the conditions under which it
can be broken have been the subject of a great deal of
theoretical work \citep{gould95,boutreux,gaudigould97,gssim}.
By contrast, despite the fact that more than a dozen microlens parallaxes 
have been measured using the Earth's motion
\citep*{al1,alcock01,Ma99, OGLE, bond01, OGLEII, O9932, jin, blackhole, smp},
the existence of a discrete microlens
parallax degeneracy for the case of an accelerated observer was not discovered 
until the past year \citep{smp},
and the fact that there is actually a four-fold discrete degeneracy
that is analogous to equation~(\ref{eqn:spacedegen}) was only just
recently recognized \citep{lmc5degen}.

	One reason for this late development is that the discrete
degeneracy is typically broken for long events, $\Omega_\oplus t_\e\ga 1$,
where $\Omega_\oplus\equiv 2\pi\,\rm yr^{-1}$.  However, this cannot be the
full explanation; \citet{smp} showed that some previously analyzed events
had unrecognized discrete degeneracies, and the first analysis of a four-fold
degeneracy came 10 years after the event was discovered.
Another reason, perhaps, is that while the four-fold degeneracy 
assumes a relatively simple form in the geocentric frame \citep{lmc5degen},
all previous parallax analyses were carried out in the heliocentric
frame, in which it is much more difficult to recognize the physical
symmetries that underlie this degeneracy.

	Whatever the exact reason for their prior neglect, it is important
now to gain a better understanding of discrete parallax degeneracies,
which are expected to be significant primarily in events that have good data
and that lie below the threshold $\Omega_\oplus t_\e \la 1$.  For cases in
which this degeneracy can be resolved by an independent measurement
of the lens-source relative proper motion (e.g., \citealt{alcock01}),
it will be possible to measure the lens mass and distance.
Moreover, \citet{sirtf} showed that the degeneracies in space-based and 
Earth-based parallax measurements are complementary, so that the full microlens
parallax can sometimes be recovered by combining these even when each measurement
is separately degenerate.  This will have practical applications
to the {\it SIM} mission, which has a major microlensing component.  Finally,
the parallax effects in the relatively short events that can give rise to a
discrete degeneracy can also be confused with some signatures of planetary
companions to the lens \citep{planet5year}.   An understanding of this
degeneracy is therefore required for a proper analysis of microlensing
planet searches.

	There are strong, although not perfect, analogies between the 
space-based discrete parallax degeneracy first analyzed by \citet{refsdal66}
and the Earth-based degeneracy.  In both cases, there is a preferred 
$(\parallel)$
axis, and the parallax component along this axis, $\pi_{\e,\parallel}$ is
hardly affected by the degeneracy.  For the space-based case, this axis
is the Earth-satellite separation vector, while for the Earth-based case
it is the direction of the Sun's apparent acceleration (in the geocentric
frame).  In both cases, there is a two-fold degeneracy that basically does not
affect the magnitude of $\bpi_\e$. For the space-based case, this is due to the
first ``$\pm$'' in equation~(\ref{eqn:deltabeta}) and corresponds to a parity
flip of the entire event, with the lens reversing the side it passes the source
as seen from both the Earth and the satellite.  For the Earth-based case, this 
two-fold degeneracy takes $u_0\rightarrow -u_0$ and so also corresponds to the
lens
reversing the side it passes the source.  Finally, in both cases, there is an 
additional two-fold degeneracy that does
affect the magnitude of $\bpi_\e$.  
For the space-based case, this is due to the second ``$\pm$'' in 
equation~(\ref{eqn:deltabeta}) and corresponds to the lens passing
on the same versus the opposite side of the source as seen from the Earth
and satellite.  For the Earth-based case, this degeneracy can be expressed 
in terms of the so-called ``jerk parallax'',
\begin{equation}
\bpi_j \equiv {4\over 3}\,{\bj\over \alpha^2 t_\e},
\label{eqn:jerkpar}
\end{equation}
where $\balpha$ is the (2-dimensional) acceleration of the Sun in the
Earth frame projected onto plane of the sky and divided by an AU, and where
$\bj\equiv d\balpha/dt$ is the jerk.   The degeneracy arises because
event geometries with the same fourth-order Taylor coefficient of the
lens-source separation,
\begin{equation}
C_4 = {\alpha^2\over 4}(\pi_\e^2 + \bpi_j\cdot\bpi_\e)
+{1\over 12}{\Omega_\oplus^2\over \alpha}u_0\pi_{\e,\perp},
\label{eqn:c4def}
\end{equation}
will, to this order, generate the same lightcurve.  
Hence, in the limit $|u_0|\ll 1$, if
$\bpi_\e$ is one solution, then $\bpi_\e'$, with
\begin{equation}
\pi_{\e,\parallel}' = \pi_{\e,\parallel},
\qquad
\pi_{\e,\perp}' = -(\pi_{\e,\perp} +\pi_{j,\perp}),
\label{eqn:bpieprime}
\end{equation}
is also a solution.

	While theoretical investigations are essential for understanding
the jerk-parallax degeneracy, it is also important to analyze individual 
degenerate events.  In particular, \citet{lmc5degen} predicted that
the phenomenology of this degeneracy should be much richer for events observed
toward the Galactic
bulge than for those seen toward the Large Magellanic Cloud (LMC)
because the bulge lies near the ecliptic while the LMC lies near the
ecliptic pole.  The jerk-parallax can be reexpressed in terms of a
projected velocity,
\begin{equation}
\tilde v_j \equiv {\au\over \pi_{j,\perp}t_\e} = 
{3\over 4}\,{\alpha^3\,\au\over\balpha\times \bj},
\label{eqn:tildevj}
\end{equation}
which (approximating the Earth's orbit as circular) can be evaluated,
\begin{equation}
\tilde v_j = -{3\over 4}{(\cos^2\psi\sin^2\beta + \sin^2\psi)^{3/2}
\over \sin\beta}\,v_\oplus.
\label{eqn:tildevj2}
\end{equation}
Here $\beta$ is the ecliptic latitude, $\psi$ is the phase of the
Earth's orbit relative to opposition, and $v_\oplus=30\,\kms$ is the
speed of the Earth.  (Note that $\balpha\times\bj$ is a signed scalar.)\ \ 
Toward the LMC (near the ecliptic pole),
$\tilde v_j \simeq 22\,\kms$, independent of the time of year.
However, toward the bulge (near the ecliptic), $\tilde v_j$
varies from a factor $\sin^2\beta$ below this value to $|\csc\beta|$ above it.
Since $\tilde v_j$ characterizes the jerk-parallax degeneracy through
equations~(\ref{eqn:bpieprime}) and (\ref{eqn:tildevj}), the bulge's
much broader range of values for this parameter implies a broader range
of phenomena.

Here we analyze MOA 2003-BLG-37, the first bulge microlensing event whose
parallax solution has four distinct minima.  We show that the analytic
formalism developed by \citet{lmc5degen} to describe the jerk parallax
applies quite well to this event.  However, the morphology of the
$\chi^2$ contours of this event is very different from that of
MACHO-LMC-5, the only other event for which the jerk-parallax degeneracy
has been detected.

\section{Data
\label{sec:data}}

MOA 2003-BLG-37 [(RA,DEC) = (18:12:2.24,-29:01:01.3), $(l,b)=(3.152,-5.656)$]
was alerted by the Microlensing Observations for Astrophysics
(MOA) collaboration on 2003 June 24 as part of its ongoing alert
program \citep{bond01}
%(HJD'=2825)
when the lens was about 1.5 Einstein radii from the source.  MOA
obtained a total of 626 observations in $I$ band using the 0.6 m
Boller \& Chivens telescope at Mt.\ John University Observatory
in New Zealand.  Of these, 359 observations were during the 2003
season and the remaining 269 were during 3 previous seasons.
The data were reduced using on-line image subtraction.

The Microlensing Follow Up Network ($\mu$FUN, \citealt{yoo})
commenced observations approximately two weeks before the peak,
which occurred on HJD$'\equiv$HJD$-2450000=2880.68$ (UT 2003 August 29.18),
using the ANDICAM camera on the 1.3m SMARTS (former 2MASS) telescope
at Cerro Tololo Interamerican Observatory at La Serena, Chile.  $\mu$FUN
obtained 116 observations in $I$ and 13 in $V$.  The photometry was carried 
out using DoPHOT \citep{dophot}.
ANDICAM is equipped with an optical/IR dichroic beam splitter, allowing 
$H$ band observations
to be taken simultaneously with each optical image.  However, the
$H$ band data show much larger scatter than the optical data, and
so their incorporation does not significantly affect the solution nor
reduce the errors.  Hence, we exclude them.

After the event peaked, we realized that it might have a measurable parallax,
and we therefore made special efforts to observe the event regularly until
late in the season so as to obtain good coverage of its falling wing.
Since the Sun's acceleration projected on the plane of
the sky reaches a maximum on about Sept 24
(roughly 0.6 Einstein radii after peak), the falling wing is the
most sensitive to parallax.  Observations continued several times per
week until the end of October at both observatories.  Reported errors at both
observatories were typically 0.01 to 0.02 mag
near baseline and smaller near the peak.

Of the 626 MOA points, four were found to be mild ($\sim 3.5\,\sigma$)
outliers and were eliminated from the analysis.  The MOA errors were
renormalized upward by a factor 1.47 in order to make their $\chi^2$ per
degree of freedom (dof) consistent with unity.  The $\mu$FUN $I$ and $V$ 
data did not contain any outliers greater than $3\,\sigma$, and their
$\chi^2$/dof were both consistent with unity, so no adjustments were 
applied.

\section{Event Characteristics
\label{sec:char}}

\begin{figure}
\plotone{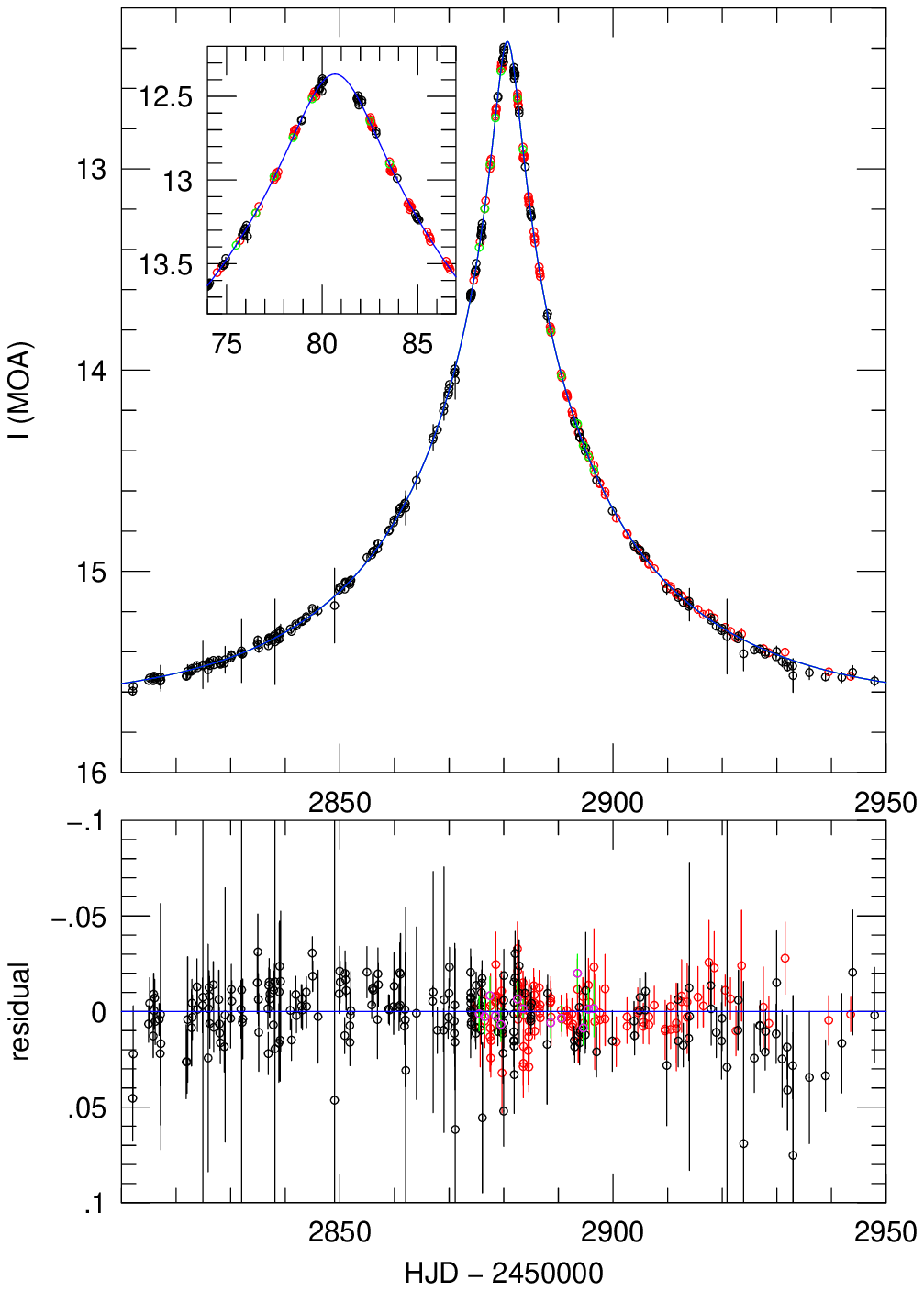}
\caption{Lightcurve of MOA 2003-BLG-37 showing data from MOA $I$ 
({\it black}), $\mu$FUN $I$ ({\it red}), and $\mu$FUN $V$ ({\it green}).
The fit does not allow for parallax.  The residuals show a very slight
asymmetry, with the data brighter than the model on the rising wing
and fainter on the falling wing.  The MOA data are shown as recorded,
and each $\mu$FUN dataset has been linearly rescaled to be consistent
with the MOA system.  The MOA scale is not rigorously calibrated but
is accurate to a few tenths of a magnitude.
\label{fig:lc}}
\end{figure}

\begin{figure}
\plotone{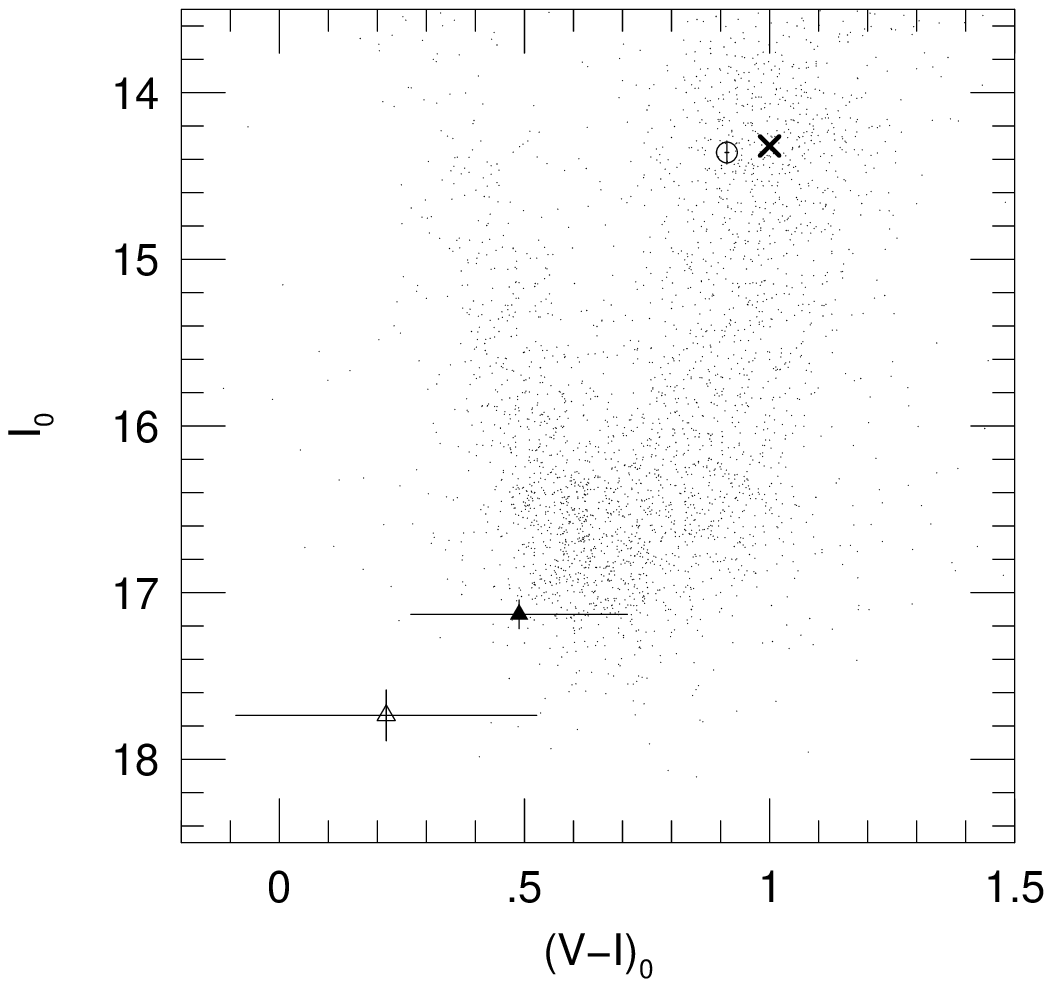}
\caption{Instrumental color-magnitude diagram of the $6'\times 6'$ field 
centered on MOA 2003-BLG-37.  The diagram has been translated so
that the centroid of the bulge clump giants ({\it cross}) is at
its known position $[I_0,(V-I)_0]_{\rm clump}=(1.00,14.32)$.
The open circle shows the color and magnitude of the (deblended)
unmagnified source as determined from the fit to the lightcurve.  
Using this diagram and the standard technique described in \citet{yoo}, 
we find that the source has an angular
radius of $\theta_* = 5.8\pm 0.5\,\muas$. The solid triangle shows the
color and magnitude of the ``hole'' in the mottled background of bulge
turnoff stars that is required to account for the negative blending
in the no-parallax model.  When parallax is included, the magnitude
of this ``hole'' ({\it open triangle}) shrinks, and is more in line
with what is expected.
\label{fig:cmd}}
\end{figure}

The lightcurve was initially fit to a point-source point-lens (PSPL)
\citet{pac86}
model, the parameters for which are shown in Table 1.  It is a relatively
high-magnification event, with maximum magnification $A_{\rm max}\sim 20$.
The data generally cover the lightcurve quite well.  See Figure~\ref{fig:lc}.
Figure~\ref{fig:cmd} is an instrumental color-magnitude diagram (CMD) of
the field, which is based on $\mu$FUN $V$ and $I$ data, that has
been translated so that the centroid of the bulge clump (marked by a cross) 
lies at its known position \citep{yoo},
$[I_0,(V-I)_0]_{\rm clump}=(14.32,1.00)$.  For stars that lie behind
the disk reddening screen, the diagram therefore shows their
dereddened color and magnitude, while foreground stars that lie within the
dust distribution (e.g., the ``reddening sequence'' toward the upper
left) appear in the diagram as brighter and bluer than they actually are.
The dereddened color and magnitude of the microlensed
source (when unmagnified) are shown as an open circle.  These
values are determined from the fits for the deblended source fluxes
$f_s$ in $V$ and $I$ from $\mu$FUN, i.e., the same data set used
to construct the CMD.  Using jackknife, we find that the statistical
errors from centroiding the clump are 0.02 mag in each axis.
The entire procedure assumes a uniform reddening
over the $6'\times 6'$ field.  In fact, we find that when
we consider stars within $100''$ of the source star (and so shrink
the sample size by 75\%), the clump centroid shifts by only 0.02 mag
in color and 0.05 mag in brightness, which are both consistent with
statistical fluctuations.  That is, there is no evidence for
differential reddening on the relevant scales.

Based on its position in the CMD, the source could be a clump giant
in the bulge but it could equally well be a disk first-ascent giant that
lies $\sim 1$ mag in the foreground.  We determine the angular size of the
source from its position on the instrumental CMD using the standard
approach as summarized by \citet{yoo}.  Very briefly, we use 
standard color-color relations to convert the observed $(V-I)_0=0.91$
to $(V-K)_0=2.06$.  Then using the dereddened $V_0=15.27$ and the
\citet{vanb} empirical color/surface-brightness relation,
we find,
\begin{equation}
\theta_* = 5.8\pm 0.5\,\muas.
\label{eqn:thetastar}
\end{equation}
The error is almost entirely due to the 8.7\% scatter in the
\citet{vanb} relation.  The next largest source of error comes
from the 0.07 mag uncertainty in the fit to the source flux, which
contributes about 3\% in quadrature to $\theta_*$.  Other sources,
such as the 0.004 mag error in the fit color and the 0.02 mag errors
in the color and magnitude of the clump centroid, contribute still
less.  The total error is 9.5\%.  (Note that this entire procedure
rests on instrumental magnitudes, so absolute calibration of the
photometry is not necessary.)

% 0.0010 720.7189
% 0.0100 720.7095
% 0.0200 719.1734
% 0.0210 719.0510
% 0.0220 718.9300
% 0.0230 720.1035
% 0.0240 719.5948
% 0.0260 719.7298
% 0.0280 720.1438
% 0.0300 721.9147
% 0.0320 722.3739
% 0.0340 723.3345
% 0.0400 725.7651
% 0.0420 727.0253
% 0.0440 728.6100
% 0.0500 736.0900

We consider finite-source effects, holding $\rho\equiv \theta_*/\theta_\e$
at various values while fitting for the other parameters.  We find a minimum
$\chi^2=751.37$ at $\rho=0.023$, but since this is only $\Delta\chi^2=1.7$
lower than at $\rho=0$ (see Table 1), we do not consider this to be
a significant detection of finite-source effects.  We are, however,
able to put an upper limit on $\rho$, and so a lower limit on 
$\theta_\e=\theta_*/\rho$ and hence on the lens-source relative proper motion
$\mu=\theta_\e/t_\e$.  We find,
\begin{equation}
\rho < 0.034,\quad
\theta_\e > 170\,\muas,\quad
\mu > 1.5\,\masyr,
\label{eqn:rhoparms}
\end{equation}
at the $2\,\sigma$ level.  (At $3\,\sigma$, $\rho<0.44$.)\ \
These constraints are of marginal interest.  The typical proper motion
expected for bulge-bulge lensing events is about $5\,\masyr$, and for
disk-bulge events it is about $8\,\masyr$.  Hence, only the low-$\mu$ part
of parameter space is eliminated.  Similarly, from equation~(\ref{eqn:piedef})
for $\theta_\e$, we obtain,
\begin{equation}
{M\over M_\odot}\,{\pi_\rel\over 25\,\muas} > 0.15.
\label{eqn:mpilimit}
\end{equation}
Hence, only the very low-mass lenses with very small lens-source separations
are eliminated.  

Microlensing flux observations $F(t)$ are always fit to the form,
\begin{equation}
F(t) = f_s A(t) + f_b,
\label{eqn:foft}
\end{equation}
where $A(t)$ is the model magnification as a function of time, $f_s$
is the baseline flux of the lensed source, and $f_b$ is the background
light within the point spread function (PSF) 
that does not participate in the microlensing event.
A somewhat unsettling feature of the PSPL model is that it has negative
blending $f_b$ in both $I$ and $V$, which is detected at the $6\,\sigma$
level and so cannot be due to ordinary statistical fluctuations.  Logically,
there are only three other possible explanations: a failure of the model to 
take account of some essential physics affecting 
the lightcurve, systematic errors 
in the photometry, or a ``negative source of light'' within the PSF.  

	The last explanation is not as absurd as it might appear.  
The dense star fields of the Galactic bulge contain a 
mottled background of upper-main-sequence and turnoff
stars that lie below the detection limit.  If
the source happens to lie in a ``hole'' in this background, the fit to
the blended flux will be negative.  By comparing the negative blend fluxes,
$f_b$, to the source fluxes $f_s$, and referencing this comparison to the CMD,
we find that the ``hole'' has  a dereddened color and magnitude of 
$(I,V-I)_{0,\rm ``hole''}\sim (17.1,0.5)$, which is about the right
color but somewhat too bright to be a turnoff star.  The position of
this ``hole'' is shown as a filled triangle in Figure~\ref{fig:cmd}.

	While systematic errors are possible, it seems unlikely that
they would afflict the independently measured $V$ and $I$ photometry
by similar amounts.  The remaining possible explanation is an inadequate model.
Since blending is even in time (about the event peak), it can easily be
confused with other even effects.  The two most obvious possibilities
are finite-source effects and parallax.  The above mentioned $1.3\,\sigma$
``detection'' of finite-source effects does indeed tend to reduce the
negative blending, but by a negligible amount.  The $\pi_{\e,\parallel}$
component of $\bpi_\e$ gives rise to a third-order term, which is
asymmetric in time and hence cannot be confused with blending.  However,
the $\pi_{\e,\perp}$ component gives rise to a fourth-order term, which 
can be confused with blending \citep{smp}.

\section{Parallax Fits
\label{sec:parallax}}

The lightcurve
shows a very minor asymmetry, being slightly brighter than the model
on the rising side and slightly fainter on the falling side.  
See Figure~\ref{fig:lc}.  This, together with the negative-blending problem 
described in \S~\ref{sec:char}, lead us to fit the lightcurve 
for parallax.  We employ the geocentric formalism of \citet{lmc5degen}.
We minimize $\chi^2$ using Newton's method, starting with the no-parallax 
solution as a seed.  We find a solution whose $\chi^2$ is lower by
approximately 32, which we regard as significant.  See Table 1.
To find the corresponding, $u_0<0$ solution predicted by \citet{smp}, 
we adopt a seed solution with $u_0\rightarrow -u_0$ and with all other
parameters the same as the $u_0>0$ solution.  We find that $\chi^2$ is
minimized extremely close to this seed.  The difference in $\chi^2$
of these two solutions is less than 1.  See Table 1.

\subsection{Jerk-Parallax Predictions
\label{sec:jerkpred}}

To navigate the jerk-parallax degeneracy, we employ the formalism of
\citet{lmc5degen}.  We first evaluate the acceleration and jerk of the
Sun projected onto the plane of the sky at time $t_0$.  We use the
method of \citet{jin} to find the path of the Sun projected on the
plane of the sky and take numerical derivatives to obtain
in (north,east) coordinates,
\begin{equation}
\balpha = (0.060,0.882)\Omega_\oplus^2,\qquad
\bj = (-0.075,0.441)\Omega_\oplus^3.
\label{eqn:alphaj}
\end{equation}
Hence, the solar acceleration points almost due east, with a position
angle of $86^\circ$ (north through east).  Since $\balpha$ determines the
parallel direction, the southern $u_0>0$ solution for $\bpi_\e$ listed
in Table 1 can be represented in the $(\parallel,\perp)$ coordinate
system as,
\begin{equation}
\bpi_\e = (\pi_{\e,\parallel},\pi_{\e,\perp}) = (-0.049,0.568),
\label{eqn:bpieeval}
\end{equation}
This explains why the parallax signature in Figure~\ref{fig:lc} is so weak.  
The component of $\bpi_\e$ parallel to the solar acceleration, which
is what gives rise to the parallax asymmetry, is extremely small.
This asymmetry arises from the third-order term in the Taylor expansion of the 
lens-source separation \citep{lmc5degen}.  Since this is the lowest order
affected by parallax, it is also the easiest to notice in
 the lightcurve \citep{gmb}.  The
perpendicular component first enters at fourth order \citep{lmc5degen}
and therefore is much harder to recognize.

 From equations~(\ref{eqn:jerkpar}), (\ref{eqn:tildevj}), and 
(\ref{eqn:alphaj}), we find
\begin{equation}
\pi_{j,\perp}= 0.274,\quad {\rm or} \quad
\tilde v_j = 147\,\kms.
\label{eqn:pijeval}
\end{equation}
Combining equations~(\ref{eqn:bpieprime}), (\ref{eqn:bpieeval}), 
and (\ref{eqn:pijeval}), and then switching to (north,east) coordinates,
we predict that the degenerate solution will lie at,
\begin{equation}
(\pi_{\e,\rm north}',\pi_{\e,\rm east}') \rightarrow (0.843,-0.009)
\qquad \rm (predicted).
\label{eqn:bpieprimepred}
\end{equation}
We begin with this as a seed, and find that $\chi^2$ is minimized at
$(\pi_{\e,\rm north}',\pi_{\e,\rm east}') = (0.788,-0.023)$, which is
quite close to the predicted value.  See Table 1.

We perform a similar exercise for the $u_0<0$ solutions.  We predict the
degenerate solution will lie at 
$(\pi_{\e,\rm north}',\pi_{\e,\rm east}') \rightarrow (0.863,-0.017)$
whereas it is actually found at
$(\pi_{\e,\rm north}',\pi_{\e,\rm east}') = (0.783,-0.012)$.
We conclude that the formalism predicts the location of the jerk-parallax
degenerate solutions quite well, at least in this case.

\begin{figure}
\plotone{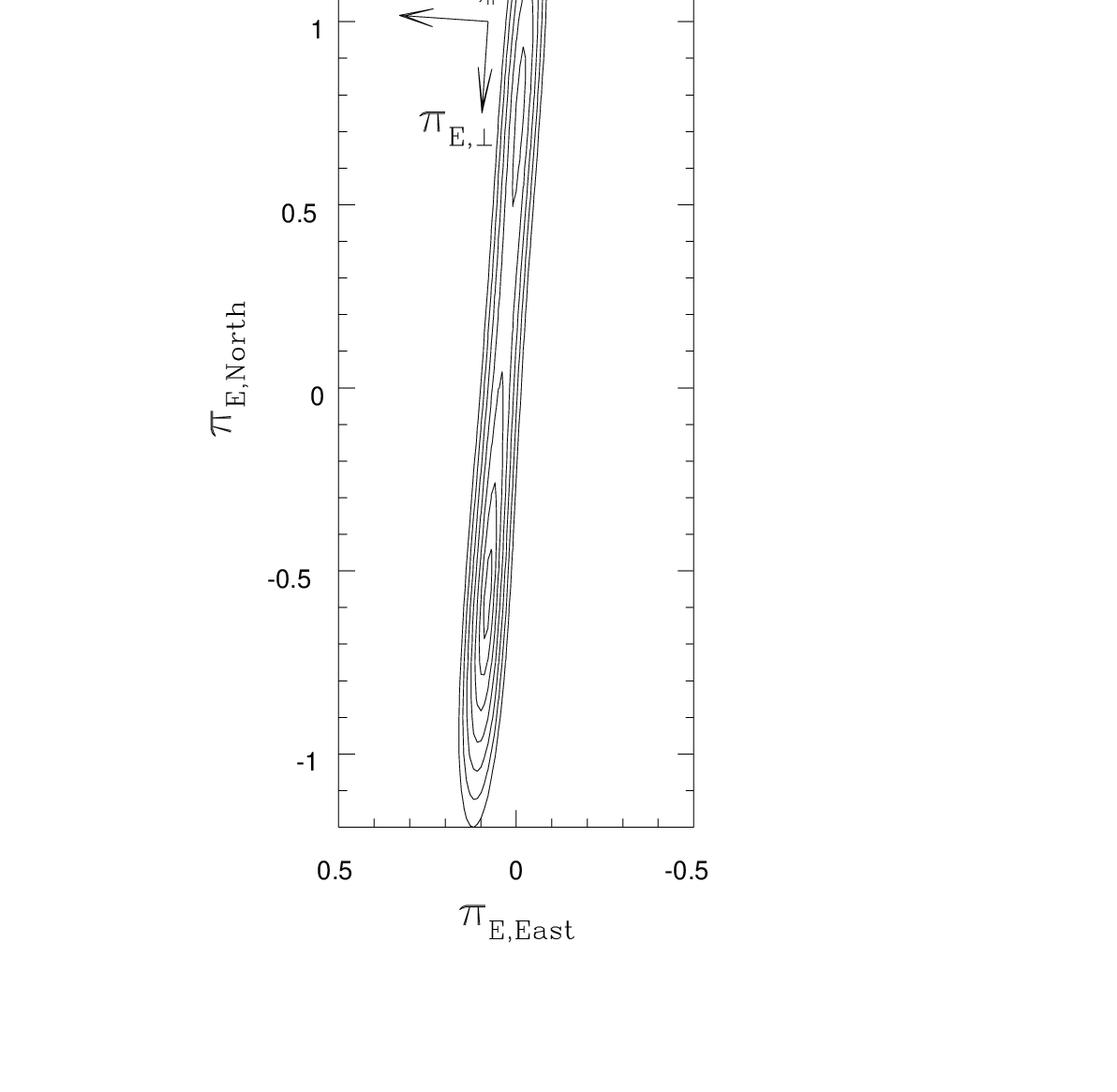}
\caption{Likelihood contours for the $u_0<0$ solutions
in the microlens parallax ($\bpi_\e$) plane shown at
$\Delta\chi^2=1$, 4, 9, 16, 25, 36, and 49 relative to the minimum.
The magnitude of $\bpi_\e$ gives the size of the Earth's orbit relative
to the Einstein radius, and its direction is that of the lens-source
relative motion.
There are two local minima, with the northern minimum higher by
$\Delta\chi^2=7.5$ and hence surrounded by the $3\,\sigma$ contour.
The two minima are embedded in contours of the continuous line-like
degeneracy, which runs perpendicular to the direction of the Earth-Sun
acceleration vector (arrows) and which reflects the parallax asymmetry.
Note that since $\pi_\e$ is dimensionless, the axes have no unit labels.
This diagram should be contrasted with figure 3 of \citet{lmc5degen}
as described in \S~\ref{sec:macholmc5} of this paper.
\label{fig:piecontours}}
\end{figure}

Some geometrical insight into this procedure can be gained from
Figure~\ref{fig:piecontours}, below.  According to 
equation~(\ref{eqn:bpieprime}), the midpoint of the two solutions
should lie at
\begin{equation}
\bpi_{\e,\rm mid} = {\bpi_\e + \bpi_\e'\over 2} = 
(\pi_{\e,\parallel},-\pi_{j_\perp}/2).
\label{eqn:pimid}
\end{equation}
The first component says that both solutions (as well as their midpoint)
should lie along a line defined by the Sun's apparent acceleration
(labeled $\pi_{\e,\parallel}$ in the figure).  This is indeed the case.
The second component says that the midpoint should lie
along this axis and $-\pi_{j,\perp}/2=-0.137$ from the ``perpendicular axis''
(which is parallel to the arrow labeled $\pi_{\e,\perp}$ and passes through
the origin).  This is also very nearly the case.

\subsection{Timescale Predictions
\label{sec:timepred}}

The jerk-parallax formalism predicts that of the remaining parameters
$(t_0,u_0,t_\e,f_s,f_b)$, all but $u_0$ and $t_\e$ should be almost the
same for all four solutions.  Moreover, $u_0$ should only change sign,
while $t_\e$ should differ among solutions by,
\begin{equation}
\Delta t_\e \simeq -{\alpha t_\e^2\over 2}
(\pi_{\e,\perp}\Delta u_0 + u_0\Delta\pi_{\e,\perp})t_\e.
\label{eqn:Deltate}
\end{equation}
Table 1 shows that these predictions are also borne out.  The parameters
$t_0$, $|u_0|$, and the three sets of $(f_s,f_b)$ are all virtually
unchanged among the four solutions.  Equation~(\ref{eqn:Deltate})
predicts $\Delta t_\e=(0.60,0.89,0.46)$ days 
when comparing, respectively, the two southern, two northern, and corresponding
southern and northern solutions.  The actual differences from Table 1 are
(0.62,0.83,0.70) days for the southern and northern pairs, and
for the average of the two south/north comparisons, respectively.

\subsection{Parallax Contours
\label{sec:piecontours}}

To further explore the degeneracy, we calculate $\chi^2$ for a grid of
parallax values and show the resulting likelihood contours in 
Figure~\ref{fig:piecontours}.  There are three features to be noted.
First, there are two local minima, which as discussed above
are related to each other almost exactly as predicted by 
equation~(\ref{eqn:bpieprime}).  Second, the northern minimum is higher
by $\Delta\chi^2=7.5$ and so is formally excluded at the $2.6\,\sigma$
level.  See also Table 1.  Third, each minimum is elongated along
the $\pi_{\e,\perp}$ axis and, indeed, is embedded in a larger continuous
degeneracy also along this axis.  That is, at the $4\,\sigma$ level, there
is a continuous set of acceptable solutions that extends almost
from $\pi_{\e,\rm north}=-1$ to $\pi_{\e,\rm north}=1$, but whose
width is less than 0.005.  The two discrete solutions form islands
within this linear degeneracy.

\section{Discussion
\label{eqn:discuss}}

\subsection{Comparison to MACHO-LMC-5
\label{sec:macholmc5}}

These three features should be contrasted to the case of MACHO-LMC-5,
for which the analogous contour diagram is shown in figure 3 of
\citet{lmc5degen}.  For that event also, the jerk-parallax formalism
accurately predicted (or rather postdicted) the location of the second
minimum, and consequently the separation vector between the two solutions
was perpendicular to the Sun-acceleration vector.  
However, in other respects the
two cases are quite different.  First, the two MACHO-LMC-5 minima had nearly
identical $\chi^2$'s.  Second, these two minima were not severely elongated.
Third, they did not lie within the continuous degeneracy.

In fact, \citet{lmc5degen} argued that there should be a hierarchy of
microlens parallax degeneracies.  When the data were adequate to solve
for the source-lens separation only to second order in time, there would
be no parallax information.  When they were adequate to solve the 
separation to third order, there would be a continuous (linear) degeneracy.
When they were adequate to solve to fourth order, the continuous degeneracy
would be broken, but a discrete degeneracy would remain.  Finally,  if higher 
order terms could be measured, the
degeneracy would be completely lifted.  This picture would seem to argue
that the discrete degeneracy should be embedded in the continuous
degeneracy as it is in the present case of MOA 2003-BLG-37.  It
is curious therefore that for MACHO-LMC-5, the degeneracy is not
embedded, even though it is more severe (in the sense the two minima
have almost identical $\chi^2$).

\subsection{Negative Blending
\label{sec:negblend}}

After including parallax in the fit, negative blending remains in both
$V$ and $I$, but it is less severe.  The dereddened color and magnitude
of the ``hole'' is,
\begin{equation}
I_{0,\rm ``hole''} = 17.7\pm 0.2,\qquad
(V-I)_{0,\rm ``hole''} = 0.2\pm 0.3,
\label{eqn:hole}
\end{equation}
where the errors are approximately independent.  The hole still has 
the right color (within errors)
to be due to fluctuations in the turnoff-star background, and its
amplitude is substantially smaller and so is consistent with an
upper main-sequence or turnoff star.  See Figure~\ref{fig:cmd}
({\it open triangle}).

\subsection{Xallarap
\label{sec:xallarap}}

As \citet{smp} point out, any lightcurve that can be fit by parallax
can also be fit by xallarap (accelerated motion of the source).
For long events with excellent data, the parallax parameters are
so well determined that the putative xallarap solution can succeed
in mimicking parallax only by having the source mirror the projected
motion of the Earth to extremely high precision.  In such cases,
one may apply Occam's razor to argue that the xallarap solution
is unlikely.  For some other events, there may be external information
that tends to corroborate the parallax interpretation, even if 
the parallax solution itself is not tightly constrained.  For example,
the lens of MACHO-LMC-5 was directly imaged which directly confirmed
the direction of motion \citep{alcock01} as well as the lens mass and distance
\citep{lmc5degen} derived from the parallax solution.

In the present case, the only auxiliary information comes from the
constraint $\theta_\e>170\muas$, which was derived from the lack of 
finite-source effects.  See equation~(\ref{eqn:rhoparms}).  If the source is
assumed to be at $D_s=8\,\kpc$, this implies that the Einstein
radius projected on the source plane is $\hat r_\e = D_s\theta_\e > 1.36\,\au$.
To mimic the effect of the Earth's acceleration, the source must 
have a projected acceleration $a_{s,\rm proj}$ such that 
$a_{s,\rm proj}/\hat r_\e = a_{\oplus,{\rm proj}}/\tilde r_\e$, where
$a_{\oplus,{\rm proj}}=\alpha\au$ is the projected acceleration of
the Earth.  That is,  $a_{s,\rm proj}=\pi_\e\alpha\hat r_\e=
\pi_\e\alpha D_s\theta_\e > 0.6 a_\oplus$, where $a_\oplus=\au\Omega_\oplus^2$
is the (full) acceleration of the Earth.  Such a source acceleration is 
of course possible.  However, if such an acceleration were induced by
a companion of mass $m$, then the velocity semi-amplitude would be at
least $v> 23\,(m/M_\odot)^{1/2}\kms$.  Hence, unless the companion
were substellar or the system were seen pole on, it would be possible to
detect such motion by radial velocity measurements.

\section{Conclusions
\label{sec:conclude}}

We have shown that MOA 2003-BLG-37 suffers from a four-fold
jerk-parallax degeneracy.  The two pairs of solutions
have microlens parallaxes of $\pi_\e=0.57$ and $\pi_\e'=0.78$.
These correspond to projected Einstein radii of
$\tilde r_\e \equiv \au/\pi_\e = 1.76\,\au$ and $\tilde r_\e'=1.28\,\au$.
The first solution is favored at the $2.6\sigma$ level.  We
were able to place only lower limits on the angular Einstein radius,
$\theta_\e>170\,\muas$.  Combining the favored parallax measurement
with this limit yield lower bounds on the mass and lens-source relative
parallax of
\begin{equation}
M>0.037\,M_\odot,\qquad
\pi_\rel > 100\,\muas,
\label{eqn:mpirel}
\end{equation}
neither of which probes a really interesting regime.

MOA 2003-BLG-37 is only the second event for which four separate
minima have been found in $\chi^2$ as a function of the vector
microlens parallax $\bpi_\e$.  The other was MACHO-LMC-5.  For
both events, the jerk-parallax formalism of \citet{lmc5degen}
accurately predicts the position of the second pair of solutions given that
the first pair has been found.  This gives hope that the additional
solution will easily be found in the future events that contain them.

On the other hand, there are substantial differences between the
morphologies of the $\chi^2$ contour diagrams for these two events
in terms of the shapes and relative depths of the minima as well as
the degree to which the minima are embedded in the lower-order
continuous degeneracy. Further understanding of these differences
is likely to require additional theoretical work as well as the
study of additional events that display the jerk-parallax degeneracy.
In particular, it would be useful to check all events with measured
parallaxes to determine which of these are affected by this
degeneracy.  The black-hole candidates of \citet{O9932} and \citet{blackhole} 
would be especially interesting in this regard.

%\begin{equation}
%\label{eqn:}
%\end{equation}

%\begin{equation}
%\label{eqn:}
%\end{equation}

\acknowledgments
We thank the referee for making numerous comments and suggestions
that significantly improved the paper.
The MOA project is supported by the Marsden Fund of
New Zealand, the Ministry of Education, Culture, Sports, Science and
Technology (MEXT) of Japan, and the Japan Society for the Promotion of
Science (JSPS).
Work at OSU was supported by grants AST 02-01266 from the NSF and
NAG 5-10678 from NASA.
Work by CH was supported by the Astrophysical Research Center
for the Structure and Evolution of the Cosmos (ARCSEC") of Korea
Science \& Engineering Foundation (KOSEF) through Science Research
Program (SRC) program.  Work by BGP was supported by Korea
Astronomy Observatory (KAO).

%\clearpage
%\input tab

\clearpage

\end{document}